# CONSOLIDATING THE INNOVATIVE CONCEPTS TOWARDS EXASCALE COMPUTING FOR CO-DESIGN OF CO-APPLICATIONS II: CO-DESIGN AUTOMATION- WORKLOAD CHARACTERIZATION

Dhanasekar[1], Anirudh Seshadri[1], Sudharshan Srinivasan[1], Suryanarayanan[2], Akash Sridhar[2]

1 Former senior research trainees (2015-18), WARFT* || 2 Former senior research trainees (2013-16), WARFT*

**Abstract**: Many-core co-design is a complex task in which application complexity design space, heterogeneous many-core architecture design space, parallel programming language design space, simulator design space and optimizer design space should get integrated through a binding process and these design spaces, an ensemble of what is called many-core co-design spaces. It is indispensable to build a co-design automation process to dominate over the co-design complexity to cut down the turnaround time. The co-design automation is frame worked to comprehend the dependencies across the many-core co-design spaces and devise the logic behind these interdependencies using a set of algorithms. The software modules of these algorithms and the rest from the many-core co-design spaces interact to crop up the power-performance optimized heterogeneous many-core architecture specific for the simultaneous execution of co applications without space-time sharing. It is essential that such co-design automation has a built-in user-customizable workload generator to benchmark the emerging many-core architecture. This customizability benefits the generation of complex workloads with the desired computation complexity, communication complexity, control flow complexity, and locality of reference, specified under a distribution and established on quantitative models. In addition, the customizable workload model aids the generation of what is called computational and communication surges. None of the current day benchmark suites encompasses applications and kernels that can match the attributes of customizable workload model proposed in this paper. Aforementioned concepts are exemplified in, the case study supported by simulation results gathered from the XYZ simulator.

# I. Introduction

The intricacies of the many-core co-design process for achieving binding across the many-core co-design spaces to comprehend power-efficient high-performance computing systems are conceptualized with great details in [9-10]. This paper unfolds a unique methodology for the co-design automation encompassing all these design spaces.

While great significance is attached to the development of power efficient high-performance computing systems specifically meant for formidable applications like climate modeling, brain modeling and computational fluid dynamics [48], it is equally essential to build a co-design automation platform to capture the complex inter and intra dependencies across the aforesaid design spaces to cut down the design cycle time. The Fig.1 portrays a high-level interaction of co-design automation process with the rest of the many-core co-design spaces. The interaction among the other design spaces is conceptualized and exemplified in the companion papers [9,10].

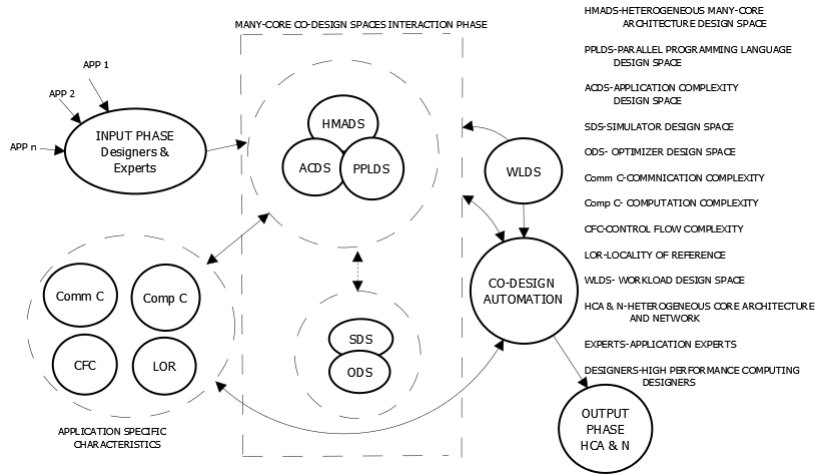

Fig.1.The figure portrays a higher level interaction of co-design automation process with the rest of the many-core co-design spaces. The attributes of the design spaces and their interactions are conceptualized and exemplified in the companion papers [9,10].

An intricate version of Fig.1 is diagrammed in section VI, characterizing the interspace dependencies.The co-design automation entails the intra and inter levels of interactions across these design spaces. To the best of the authors' knowledge, there are no published research papers on co-design automation built on the concepts deliberated in the companion papers [9,10] and this present paper.

It is imperative that such many-core co-design spaces powered by an automation process, have a built-in user-customizable workload generation mechanism to benchmark the emerging target architecture.An ingenious concept on workload model which is comprehensively customizable is proposed. The customizability here refers to the fact that, one can generate workloads of specific computational, communication and control complexities, given a distribution. The locality of reference a crucial aspect of an application or a program based on which the cache performance lies. The workload customizability allows random variations in the locality of reference both space and time. The workload model has entrenched on graph theoretic principles, allowing for the quantification of these complexities. This quantification is of prime concern in designing appropriate workloads to benchmark heterogeneous many-core architecture either comprehensively or its individual components like cache hierarchy, main memory, network, scheduler and complex functional units. Anotherprincipalfacet of this workload model is the provision for generating computational, communicational and control flow surges, and this will help focuses on the behavior of heterogeneous many-core architectures during either the cold start or during any time period of its execution.

Benchmarking results of heterogeneous many-core architecture, originated from the co-design automation, for the target co applications, delineated in the case study,are analyzed and presented in section IX, These benchmark results include overall comprehensive performance evaluation of the target heterogeneous many-core architecture, individual benchmarking of cache, networks, performance of functional units, main memory bank allocation and the scheduler.

The customizable workload model is such that, it can be utilized for application cloning purposes also. Application cloning is a significant concept for designing either a new high-performance computing system or evaluating the suitability of existing such systems for classified applications and also to overcome any IP violations.The Fig.2 illustrates a higher level abstraction of the proposed customizable workload generation model.The customer using the Application Complexity Modeling tool(ACM ) extracts the computation cum communication graph of the applications characterizing the complexities namely C1, C2, C3, and C4.The workload design engine generates a graph-theoretic based workload model with similar complexities but giving totally a different computation cum communication graph with the help of the algorithm bank (ALGOBANK) which has a repertoire numeric, semi-numeric, non-numeric algorithms and general purpose operations along with their respective complexities measures. Thus, the application gets completely hidden. With the workload design engine, any numbers of complex benchmark suits can be generated as per the specified distributions for the respective complexities. Thus, the co-design automation and customizable workload model form the central theme of this paper.

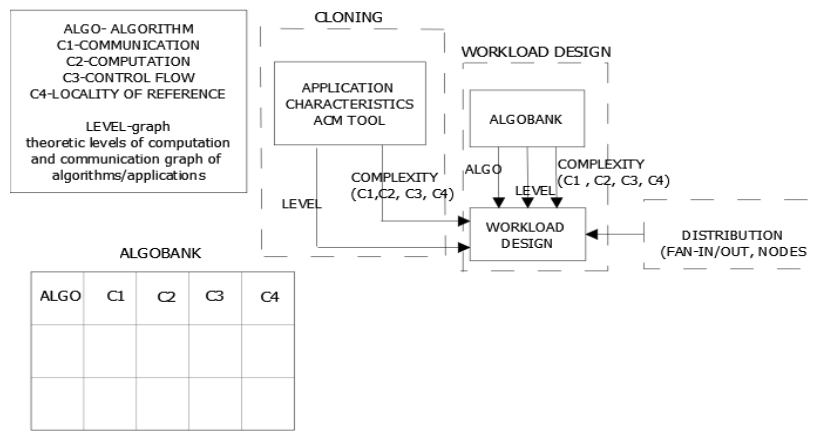

Fig.2.Higher level abstraction of customizable workload generation: Benchmarking and Cloning.

## Organization of this paper:

From the perspective of heterogeneous many-core architecture, section-confers highly correlated published research papers on workloads and benchmark suites. This section explains benchmark suites like SPEC[], LINPACK[], IBS[] ,SPLASH2[], PARCEC[],Rodinia[] with their merits and demerits.A generic model of the customizable workload, established on the principles of graph theoretic approaches, associated with computational complexity, communication complexity, control flow complexity and locality of reference, is exemplified, in section III. In section IV the implementation aspects of user customizable workload model are detailed.

The related research papers on design automation are presented in section V. The interaction among the many-core-design spaces (refer Fig.1) during the co-design automation is brought out in section VI. The section VII presents that, how the different types of cores and their respective optimal count, specifically tuned to the simultaneous execution of multiple applications without space-time sharing, are created from the isolated components like algorithm level functional units, scalars, registers etc. that initially emerge out from the application complexity design space based on the inputs, namely the multiple applications. The design of different algorithms meant for the co-design automation is discussed in this section.Co-design of high-performance computing systems [9] for simultaneous execution of multiple applications without space-time sharing[smapp], leads to power-performance efficient heterogeneous many-core architecture and efficient heterogeneous inter-core heterogeneous mesh network. Design of such heterogeneous network is detailed in section viii.

A case study on the co-design automation is actualized in section IX for designing a many-core architecture along with the correlated ISA, specific to a set of multiple applications for simultaneous execution without space-time sharing. As an outgrowth of the co-design automation, clones are created for these set of multiple applications and it is shown that they get well correlated with the complexities (computation, communication, and control) of the multiple applications.

## II.Heterogeneous many-core architecture benchmark suites: relevant papers

There are two classes of workloads. The first one is for benchmarking high-performance computing systems and the second type is for the application (classified and IP based) cloning [25-28] to decide on a suitable commercially available high-performance computing system.

The popular benchmark workloads are LINPACK [25], IBS [29] and SPEC [26]. The SPEC benchmark includes a set of applications and kernels. These applications are based on the different class of algorithms used in the domain of science and engineering. The spec includes several suites.The bzip2 is extensively used for file compression and has nine levels of compression stack involving Run-Length encoding, Burrows-Wheeler transform, Move to Front transform, Run-Length encoding on MTF, Huffman coding, Selection between multiple Huffman tables, Unary base 1 encoding, Delta encoding, Sparse bit array [30]. The single depot vehicular scheduling makes use of combinatorial optimization applied to graph-theoretic algorithms like traveling salesperson. Video Compression involves image compression algorithms like singular value decomposition, JPEG [31]. Astar makes use of shortest path finding algorithms with imposed constraints like move speed and passable/non-passable terrains. The Linpack benchmark [33] involves solving a dense system of linear equations based on matrix algorithms like LU decomposition and singular value decomposition. The Isolation Benchmark Suite has the following benchmarks, CPU intensive tests, memory intensive, a fork bomb, disk I/O intensive, the network transmit intensive and network receive intensive [34]. The "Memory stress test" adopts continuous allocation of memory using functions like calloc in Linux [29] , whereas the "I/O stress test" does continuous read-write disk operations involving heavy data movement.

Apart from SPEC, LINPACK and IBS, there are other benchmark suites which include complex workloads either in the form of real applications or application kernels. Among these benchmark suites, the notables are SPLASH2 [splash] , PARSEC[ parsec] and Rodinia[45]. In SPLASH2, the application/ workload suites are parallel programs chosen based on the characterization along different directions namely speedup, load balancing, working sets, and communication to computation ratio and issues related to spatial locality. For more details on these refer [splash]. In all, there are twelve workloads covering a wide range of applications and computational kernels. The PARSEC benchmark workloads are relatively more advanced in comparison with the splash2. According to PARSEC benchmark workload description, diversity, multi-threaded applications employing state of art techniques are good enough to support research and emerging technology. There are in all twelve workloads described in parsec [parsec]. The characterization of parsec benchmark workloads is along the similar lines as presented in splash2 namely parallelization, working sets, locality, communication to computation ratio and off-chip traffic. The parsec benchmark workloads provide parallel programs for the evaluation of chip multiprocessors.

The "Rodinia" benchmark suite introduced in [45] is meant for performance evaluation of highly specialized computing systems which includes GPUs, accelerators, FPGAs and STI Cells [46]. The importance of this benchmark suite is, that it works on various types of workloads patterns, parallelism and data sharing. This suite includes applications like dynamic programming, dense linear algebra, MapReduce, graph traversal and more to be expected in the future as given in table 1 of [45]. According to [Rodinia] the algorithms involved in splash2 workloads are for homogeneous systems and have become obsolete and they lack software pipelining. GPUs and accelerators are not supported in SPLASH and PARSEC, unlike the Rodinia. Also, the PARSEC applications support an only modest number of cores and hence cannot be ported to many-core systems such as GPUs. However, in building power-performance efficient heterogeneous many-core architectures of the future, minimizing the on-chip data moment becomes important [exa-1, exa-2]. The number of on-chip accelerators [Intel,nvidia] tends to drastically increase the on-chip data moment. The Rodinia benchmark suite includes applications and computational kernels like leukocyte tracking, back propagation, k-means, breadth-first search etc. The selection of workloads in Rodinia is based on the set of parameters like any other benchmark suites.

A workload model generation specific to applications for exploring embedded system-level design has been proposed in [35]. Here the generation methodology is based on modified GCC compiler to capture the application characteristics on a realistic basis. An improvement over the workload accuracy presented in [35] by including the workload extraction of precompiled libraries and tracking the control flow more accurately [36]. To overcome the time complexity of simulating large-scale applications to design heterogeneous many-core architecture, proxy applications and proxy architectures are proposed [37-38]. Such an approach is adopted in [39] to synthesize workload for branch prediction and memory pattern (critical aspects of an application) for benchmark suites Barnes, Cholesky, Ocean-C, FFT, and LUD. These workloads, which are thread based have reduced time complexity yet matching accuracy (+- o.17 % to +- 11.7%) with regarding CPI, Cache hit rates and branch prediction compared with the above-mentioned benchmarks.

However, to maintain very high accuracy for achieving optimal power performance scalability, not to compromise on time complexity, RTL workloads are generated and executed on FPGAs to reduce time complexity[40]. Another approach is to partition the application workload graph generation in a parallel environment

# III. GENERIC MODEL FOR CUSTOMIZABLE WORKLOAD

A deeper analysis of the mechanism involved in selecting benchmark workloads (eg.Splash2 Vs ParsecVsRodinio) reveals that their lifespan is limited o a time window[rodinia] during which the technology-driven heterogeneous many-core architecture complexities to very very rapidly and with a lot more challenging applications coming forth. The most fundamental issue is that the workloads need to be built on quantitative modeling of computational complexity, communicational complexity, control flow complexity and locality of reference. Further to this, quantification of the workload characteristics and customizability are other important factors to be considered such that scaling up the workload characteristics along with the technology time frame and forthcoming applications in the domains of science, technology, and engineering will be possible. Such workload model is useful either for general purpose or application-specific issues. In a generic workload model using which a user should be able to customize computation, communication, control complexities and the locality of references and design either a comprehensive benchmark to evaluate the overall performance or to benchmark the individual components to suit one's needs will be the ultimate. For example, the user can customize the locality of reference in such a way as to leverage the cache performance, customizing the communication complexity for surge variations in order to track the network response.

Breaking the convention: An all-embracing workload model
 There are several benchmarks [42-44] to evaluate the performance of NOC, Cache, Scheduler and functional units, overall performance but they are not based on generic customizable workload model which is more flexible to generate complex benchmark suites. To develop such a generic workload model, the computation, communication, control complexities and the locality of reference needs to be effectively quantified. Though these complexities are the major constituents of an application the control flow complexity and the locality of reference play a major part in characterizing the applications.

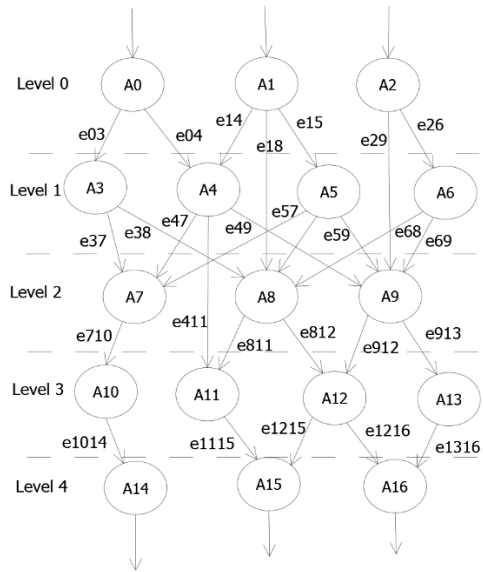

Fig.3: Weighted graph Theoretic model of a complex workload.

A graph-theoretic based workload model is given below, in Fig.3. This graph-theoretic model is based on the methodology presented in the upcoming section. The edge weights are a function of the measure of data in bytes that gets communicated from one node to the other and the frequency of communication. The node weights are either a numeric, semi-numeric or non-numeric algorithm and also, general-purpose operations. The algorithms (the weight of a node shown in Fig.3), in-degree and out-degree of the nodes, nodes per level, the total number of nodes across all the levels and the number of levels are decided based on the specified distributions of these complexities.

The most striking advantage of the proposed generic model of the workload, which helps achieve customizability as explained in the next section IV is application cloning, besides comprehensive workload generation for benchmarking various classes of high-performance computing system. In several instances, the user cannot part with the crucial details of the application to avoid IP violation and any government regulation. Using the ACM tool presented in the next section user can just generate the application computation complexity table and the application communication complexity table (examples are for complex workloads in next section IV) in which the actual application is completely hidden. The user, after extracting the computation and communication complexity using the ACM tool, can generate the corresponding workload using Customizable Workload Generation Engine (CWG) presented in the next section. In general, the grand challenge applications are very complex, hence the associated communication graph and complete cloning can be a difficult task. The ALLDE (All-Embracing Execution) [9] demands heterogeneous core architecture of MIMD class [41], for which the approach prescribed in [39] is not completely suitable.

Characteristics of an all-embracing workload model:

Both in the introduction section and in this section, the essentials of a workload model in the context of ever changing technology driven heterogeneous many-core architecture are brought out lucidly. The all-embracing model includes the following.

**Modeling Computation Complexity**

The computational complexity of an algorithm is independent of the architecture whereas the execution time may vary depending on the architectural characteristics whereas the communication complexity varies considerably upon parallelizing. The conventional computational model is adopted for workload generation. [knuth]

**Modeling Communication Complexity**

In this subsection, a quantitative model for communication complexity of an application is presented. As an example, consider the DAG shown in figure 3 in which weights of the nodes are some algorithms and edges are the links establishing the communications across the hyper nodes.

**Communication Structure Complexity:**

In this subsection, a quantitative model for communication complexity of an application is presented. As an example consider the DAG shown in figure 3 in which weights of the nodes are some algorithms and edges are the links establishing the communications across the hyper nodes.(communication graph of mst,tsp and lud and sigma model(more examples)) Building the design space for Application Complexity Modeling In the introduction section, the importance of modeling the application complexity is discussed to fix the characteristics of heterogeneous core architectures, type of functional units, different cache levels and the interconnect network for achieving binding. In general, applications encompass different class of algorithms apart from general purpose operations. This means the computational and communicational complexities of various classes of algorithms need to be analyzed and modeled. The various models presented constitute the design space meant for application complexity modeling. Communication models have been proposed with regard to establishing either randomized protocols, nondeterministic protocols and average-case protocols, with regard to communication across processes, and it is always the lower bound that is presented. However no communication model is ther for an algorithm [1]. Quantifying the communication complexity of an algorithm, is a function of the size of data(data set size) communicated between computing vertices present in the communication graph and the level of dependency involved across these data set flow along the hyper edges linking computing vertices. Let (V1, V3) be adjacent hyper vertices at levels i and k re- spectively, the depth index of the hyper edge between two adjacent( V1, V3 ) hyper vertices is given by —i k—, the ad- jacency level is non zero and positive. In general, the depth index of the hyper edge of the communication graph = —i k— where i,k is 1, 2, n where n, the number of levels is positive. The depth indices give the dependency between hyper graph vertices in the hyper graph workload. The hyper edge weight between adjacent vertices p and q $e_{pq}$ is defined as $e_{pq} = d_{pq}D_{pq}$ (1) where $D_{pq}$ is the data set size being transfered across adjacent hy- per vertices p and q. $d_{pq}$ = —i k— (2) where, p,q are adjacent hyper vertices present at level i and level k respectively.

WARFT – Waran Research FoundaTion, 45-B, Mahadevan street,West Mambalam, Chennai,India.
Director/ Founder: Prof. N.Venkateswaran

The communication complexity of the entire hyper graph is based on two important measures in every hyper vertex. The internal communication complexity(local) The external communication complexity(global) Every hyper vertex in the hyper graph workload has two communication complexities associated. The one being the communication complexity of the algorithm inside the hyper vertex. This is called the internal complexity characterized by the algorithm, and is modeled by the communication complexity taking either fan in or fan out into account. The second being the external communication complexity asso- ciated with the hyper vertex due to the fan in or fan out. Now the communication model of the entire hyper graph is given as the vector of elements where each element of the vector represents the sum of internal complexities(in terms of either fan in or fan out) and the corresponding external communication complexities (in terms of either fan in or fan out) of each hyper vertex. Example: With reference to figure 1. The internal and external communication complexity of hyper vertex A5 is calculated as follows:

### 1.4 External Complexity:

Complexity considering fan in alone : Let $D_{15}, D_{25}, D_{05}, D_{35}$ be the data set size of incoming edges, the corresponding depth index $d_{iq}$ across the adjacent hyper vertices(algorithms) as defined perviously, $d_{15}$ across A1 and A5 is one, $d_{25}$ across A2 and A5 is one, $d_{35}$ across A3 and A5 is one, $d_{05}$ across A0 and A5 is two. $CEF_{in,i}$ be the external communication complexity of the hyper vertex i with respect to fan in. depth index for edges: $d_{15}(acrossA1A5) = —1\ 2— 1$ $d_{25}(acrossA2A5) = —1\ 2— 1$ and similarly for $d_{35}$ and $d_{05}$ edge weight: $e_{15} = D_{15}\ d_{15}$ $e_{25} = D_{25}\ d_{25}$ and similarly for $e_{35}$ and $e_{05}$ $CEF_{in,5} = [D_{15}, D_{25}, D_{35}, D_{05}]\ [d_{15}, d_{25}, d_{35}, d_{05}]^T$ (3)
Complexity considering fan out alone : Let $D_{58}, D_{59}$ be the data set size of the outgoing edges, the corresponding depth index $d_{op}$ across the adjacent hyper vertices(algorithms) as per the definition given above, $d_{58}$ between A5 and A8 is one and $d_{59}$ between A5 and A9 is one. $CEF_{out,i}$ be the communicational complexity of the hyper vertex i with respect to fan out. depth index for edges equations edge weight: $e_{58} = D_{58}\ d_{58}$ $e_{59} = D_{59}\ d_{59}$ $CEF_{out,5} = [D_{58}, D_{59}]\ [d_{58}, d_{59}]^T$ (4)

### 1.5 Internal Complexity:

Internal communication complexity of the hyper vertex A5 is the communication complexity due to the fanins or fanouts present inside the hyper vertex. Let $CI_5$ represent the over all internal communication complexity, which is the commu- nication complexity due to the fanins($CIF_{in,5}$) or the com- munication complexity due to the fanouts($CIF_{out,5}$). $CI_{,5} = CIF_{in,5}$ (or) $CI_{,5} = CIF_{out,5}$.

Communication complexity of A5 : $CE_{,5} = e_{15} + e_{25} + e_{35} + e_{05} + e_{58} + e_{59}$ (or) $CE_{,5} = CEF_{in,5} + CEF_{out,5}$ $C_5 = CE_{,5} + CI_{,5}$ Now the communication complexity of the entire hyper graph is defined by the vector of C1, C2, C3, . . . Cn where Cn is the communication complexity of individual computing hy- per vertex taking either fan in or fan out. Figure 2: A sub graph of the hyper graph given in figure 1 Extracting the communication complexity of an algorithm[] The communication complexity of the graph is either the sum of its fanins edges or the fan outs edges. While the communication complexity of a single vertex alone(hyper vertex or vertex) is the sum of both fanins and fanouts components. The vector length of the models are bound to be very huge and this gives a greater insight in generating highly complex workloads. However these vector complexity measure can also be given under desired distribution. Accordingl one can specify distribution measure for communication and computation complexity in the respective bands of varying levels assuming a hypothetical very large hyper graph workload, these distribution can be varied across different bands. However the solution hyper graph workload model will be given in terms of large vectors for different bands cor- responding to the given distribution. In this paper we have considered only the vector length into account in generating the hyper graph workload model.

The relative intensity of communication complexity is given in terms of following:
CASE I : Communication complexity is intense: both depth index and data set size are large.
CASE II : Communication complexity is Medium: depth index is high and data set size are small.
CASE III: Communication complexity is Medium: depth index is low and data set size are large.
CASE IV : Communication complexity is low: both depth index and data set size are small.
With respect to high performance computing system design, it is a necessity to analyze the computational structures involved at various phases of the application and more importantly the interaction or the dependency across these computational structures. This analysis helps to understand the computation complexity and the communication complexity, which is nothing but the dependency across various computation structures.

WARFT – Waran Research FoundaTion, 45-B, Mahadevan street,West Mambalam, Chennai,India.
Director/ Founder: Prof. N.Venkateswaran

Control flow Complexity:

Based on the conditional statements, the execution flow will take one of the fan-out paths 1 to n. Refer figure 5. The control flow model is defined as a probability vector which decides the path of execution flow. The control flow vector {p1, p2, p3, . . .pn} where n is the number of fan out greater than 1. where, pi is the probability associated with the fan out path,pi 6= pj∀n> 1, pi = pj∀n (fanout) Execution flow follows the path of highest probability. These probabilities are generated under normal distribution lying in the set {0, 1}. Inclusion of this control flow model in the hyper graph describing the workload is shown in figure 5.

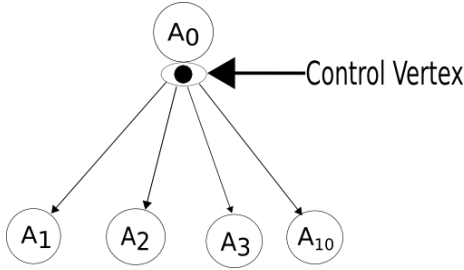

Fig.5 Control Flow Complexity model showing the probability associated with each of the control path.

Locality of reference

In order to benchmark performance of multi-level cache architectures, which is greatly affected by mapping heuristics, locality of reference(spatio-temporal) of a workload is probabilistically varied. Unconditional/Conditional loops within a workload are vital in affecting spatio-temporal locality of reference within a large class of workloads. To introduce drastic variation in locality of reference, the loop(/array) indices cannot be deterministic values but should be specified under an appropriate distribution. This is highlighted in the following loop example:

Let 'a' be the starting address defined under a random distribution 'b' be the varying incremental address under a random distribution 'c' be the ending address also under a random distribution.

$Z_a$, $Z_b$ be the indexed variables
for ( a; (a + b < c)||(loop count < condition); b )
do
$Z_a$ : f1(a, b);
$Z_b$ : f2(a, b);
loop count + +;
End

In the above loop, the base index and the increment address is specified under a distribution. This will lead to random memory addressing. The indexed variables $Z_i$corresponding to a set of expressions such that the variable are changed under the same distribution as the base index. To make things more complex, the function f1 ,f2 ,f3 can be made a function of loop indices,loop increment.

The user can customize the loop models by randomly varying thelocality of reference and include this model as an algorithm within the ALGOBANK. When these loop models become hyper-vertices in the hyper graph workload, its communication complexity of these hyper-vertices is calculated.

# IV. User customizable workload: implementation

A complex workload has to be designed using the (C3L) various models, computation complexity model, communication complexity model, Control flow complexity model and locality of reference model. Such a workload is very comprehensive due to the fact that the workload generation can be customized to benchmark individual architectural components. As special cases, specific workloads with computational surge and communicational surge can be generated to stress the high-performance computing system beyond its limit of computational elasticity (beyond the limit, the system shows odd behavior like a hang). This comprehensiveness and customizability of the workload is explained in the next section on results and analysis.

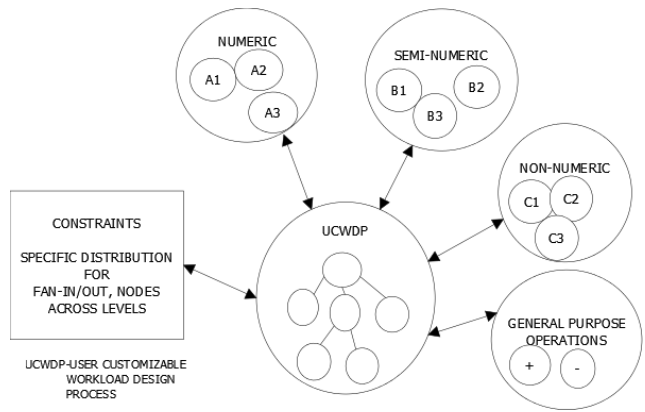

Fig.6. Graph theoretic based User Customizable Workload Design process (Refer Fig 3). The As, Bs, and the Cs are numeric, semi-numeric and non-numeric algorithms including general purpose operations. The constraints are the in degree and the out degree (fan-in and fan-out) of the nodes (of the graphic theoretic model of the of the workload) at the inter and intra levels, under user specified distributions.

the algorithm for generating graph theoretic workload shown below

*Weight (node)- Set of algorithms*

*Weight(nodepair)- algorithm*

*Edge(nodepair)- datasetsizebetweenthenodepair * (depth(node1) – depth(node2))*

*Communicationcomplexity(bensim) - Sum(Comm(i,j))*

*Computationcomplexity(bensim) -Sum(Comp(i,j)*

Complex multiple workloads chosen for ALLDEare described in Fig.8 to Fig 9.The Fig.7shows a schematic of the process.

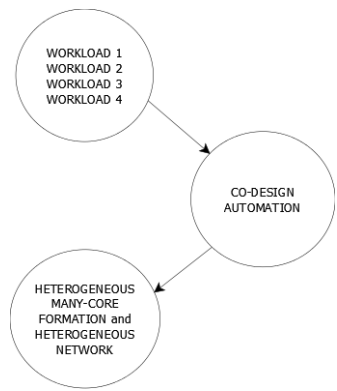

Fig.7.

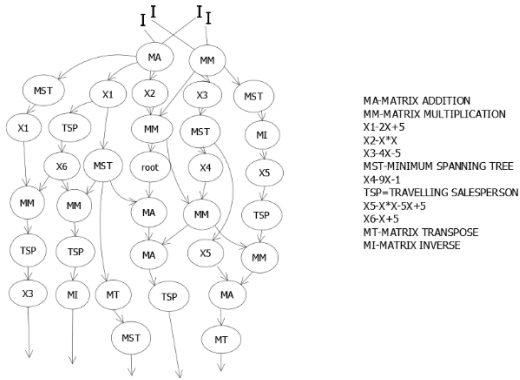

Fig.8. Matrix, Graph and general purpose based workload 1

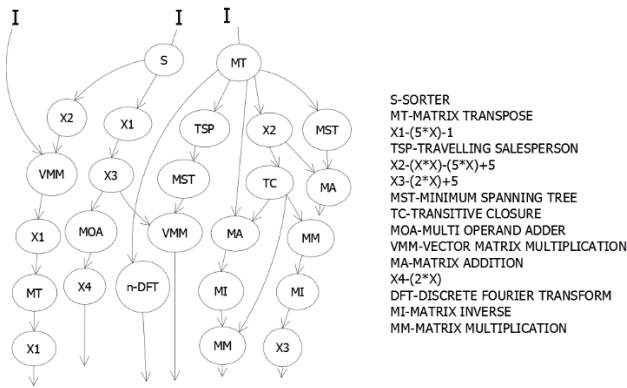

Fig.9. Matrix, Graph and general purpose based workload 2

**V. Design Automation: relevant papers**

One of the earliest research papers on design automation dealing with both architecture and application is by William Rosenbluth [1], which describes a design automation methodology for system architecture with CAD workflow using LSI components of those days. Theoretically the problems of design automation are NP-Hard, for example, partitioning, module selection, placement, fault detection and wiring in VLSI circuits [2]. With the advancement of nanotechnology and the evolution of heterogeneous many-core architecture, techniques for automatic generation of application specific multiprocessor acquired greater emphasis [3]. In [3], a design flow for such an automatic generation process inclusive of the communication co-processors optimized to the applications is portrayed. However, this design flow is indifferent to power-performance optimization, the main focus being achieving shorter design cycle. Interesting research works, on heterogeneous many-core architecture using only single ISA, heterogeneous many-core optimization for chip multiprocessor and optimal design space exploration have been reported by Rakesh Kumar et.al [4-7]. There are fundamental differences between the research works of [4-7], with the work presented in this paper and the companion papers [9, 10], concerning the design of heterogeneous multi-core architecture. The concepts devised in [4-7] is on the exploration of a single design space (heterogeneous multi-core architecture design space) for optimal (power-area) solution, whereas this paper is on co-design automation, for exploring the many-core-co-design-spaces, to build a power-performance optimal heterogeneous many-core architecture and this is the fundamental difference. Further the network architecture, an important component, particularly for heterogeneous multi core processor design has not been contemplated in [4-7]. Often, there is a strong need to explore huge design space to arrive at power-performance efficient heterogeneous core architectures. There are number of design space exploration tools like [7] to meet the demands of applications pertinent to power-performance and chip area efficiency and to reduce the design cycle time. Here software simulation and the search algorithms are the essentials. To achieve cycle accuracy, RTL simulation is carried out which is time consuming by the way, while software simulation lacks accuracy. Hence FPGAs are used to perform hardware emulation which provides cycle accurate results with reduced simulation time [16].To speed up the software simulation and the associated search algorithms for exploring the design space, heterogeneous many-core architecture simulator specific to applications are developed [17].

# VI. MANY-CORE-CO-DESIGN AUTOMATION: ZOOMING IN TO THE DESIGN SPACE PROCESSES

There are several research projects reports stressing the need for evolving a co-design methodology for designing future high performance heterogeneousmany-core computing systems apart from number of research papers[] on this.Though these project reports and research papers establish the strong necessity to pursue and evolve a co-design methodology, obviously stressing more on technology, but none of them provide a comprehensive solution

The two companion papers [9,10]  and this paper together attempt to provide a comprehensive solutionfor designing heterogeneous many-core architecture backed up by extensive simulation results [10]. The paper [9] focuses on the binding of many-core co-design spaces, whereas [10] deals with the simulator and optimization design spaces. This paper unfolds a unique methodology for the co-design automation encompassing all the design spaces probably for the first time.

The complexity of the co-design process is such, that it demands a carefully thoughtautomation technique to design power efficient high performance computing systems. Co-design automation is a complex taskof handling the integration of different design spaces about which an higher level abstraction is illustrated in Fig.1 and detailed further in Fig.11 and  Fig.12.

An interesting and inherent aspect of this co-design automation, portrayed in Fig.11 is that all the core and the uncore components including the heterogeneous inter-core networks  and also the  workload mapping  are evolved accounting for their dependencies specific to ALLDE and the parallel programming language.

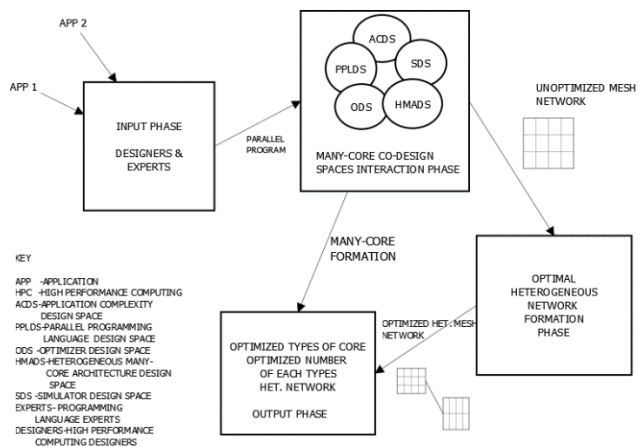

Fig.11. A schematic of Co-design automation process

The algorithms meant for the design automation need to be extremely clever to arrive at the desired solution while satisfying the number of constraints imposed on it during the exploration of the many-core co-design spaces. Efficient process flow for co-design automation needs to be fixed carefully, bearing in mind the dependencies across all these design spaces.

## The phases of co-design automation:

The co-design automation flow is depicted in Fig.11 and 12,showing the dependencies across many-core co-design spaces that are conceptualized in [9] .The input phase,  many -core co-design spaceinteraction phase, the results and

WARFT – Waran Research FoundaTion, 45-B, Mahadevan street,West Mambalam, Chennai,India.
Director/ Founder: Prof. N.Venkateswaran

analysis phase and the output phase(leading to the target architecture) together enable to resolve the types of cores and their respective count. Several algorithms involved for the same are discussed in the subsequent section.The massive complexity of co-design automation is explicit in Fig.12. In depth discussion and analysis among the architecture design experts, application experts, optimization and algorithm experts, parallel programming language experts and simulation experts are indispensable (meant for simultaneous execution of multiple applications without space time sharing), prior to the start of the co-design automation process.

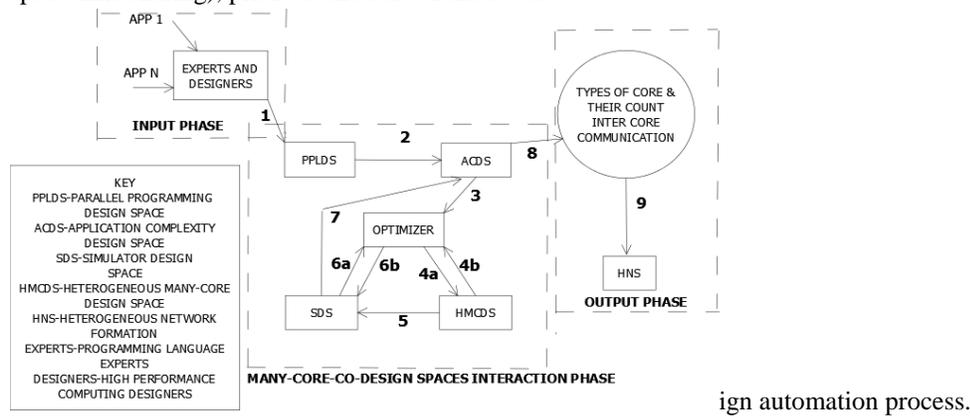

**Fig.12**.-Many-core co-design spaces interaction in co-design automation

The interaction among the design spaces as enumerated in the Fig-12 is explained below

1. Gives the various parallel programming language constructs at applications' level as resolved by the designers and experts.
2. The parallel programs for the set of applications devised by the experts.
3. The different class, numeric, semi-numeric, non-numeric and general purpose functional units (Algorithm level functional units/scalar units. This is exemplified in [9] and their intercommunication pattern.
4. The class of functional units and their communication pattern.
5. The latency, pipelining cycles of the respective function units buffers, from the heterogeneous many-core design space [9].
6. a) The resource utilization statistics functional units' connectivity matrix from the simulator.This is explained in [10].
    b) The optimized count of individual functional units, cache sizes etc.
7. The core formed from the optimized functional units and their connectivity is being fed into comparative analysis.
8. Optimized types of core type and their respective count.
9. optimized count of functional units and their inter-communication pattern after several optimization cycles from the optimizer. The optimizer design space is exemplified in [10].
10. From application complexity design space, heterogeneous network architecture is evolved.

# VII.Many-Core Formation: core types and respective counts

This section is on the design of different algorithms responsible for the development of co-design automation that are conscious of the dependencies across the design spaces portrayed in Fig.12. These algorithms follow in the upcoming subsections.

## Input to the co-design automation:

Higher level assemblylike languageprograms written by experts for the set of applications (using the computation and control constructs deduced from the parallel programming design space, specific to the application set[9]).

### Algo1: Traversal(scanning)algorithm [ ]

Input: - Higher level assembly like program code of multiple applicationswritten by the experts.(using the constructs specific to control and computational structures deduced from the parallel programming design space)

Output: -Generates graph theoretic models of computation complexity and communicationcomplexity from the higher level assembly like program code of multiple applications. Also delivers the distinct count of disparate types of functional units at each graph theoretic levels.

### *Algo1 Steps:*

1. Scans the higher level assembly like program code of multiple applications written by the experts
2. Determines the divergent computational and the control constructs.
3. Also further extractstheir inter-communication pattern,the worst case(considering the run time conditions).

## Algo2: Extracting the characteristics of the multiple applications.

Input: Inter-communication pattern of functional units obtained from Algo1.

Output: The mean and variance of the individual functional units with their inter-communication pattern. (This goes to optimizer design space)

ALGO 2: Steps:

1. Generates the quantitative model of the communication cum computation complexity graph associated with the functional units (chosen from the heterogeneous many-core architecture design space specific to the computational constructs of multiple applications).The communication cum computation complexity model is presented in sections 3 and 4 above.(include the respective fig nos)

2. Uses the variation of the individual count of different types of functional units present across the levels of communication cum computation complexity graph to arrive at the mean and variance of the individual functional units.

3.eGnerates two tables which gives graph theoretic level wise the computation complexity and the communication complexity including the count of different types of functional units at each graph theoretic level.

## Algo 3: Clustering algorithm.

Here a cluster corresponds to a small set of heterogeneous functional units which constitute a type of core.

Input:- From Algo2mean values of divergent functional units and their inter-communication pattern.

Output:-Ideal countof core types.

### *Algo 3 Steps*

1. The number of core types (desired) is set..
2. The ALFU's are then inserted into core type one by one based on the affinity (amount of bytes)
3. A list of core types with ALFUs are found.

## Algo 4: Estimating inter core communication

Input:- Inter core communication pattern
Output:-Adjacency matrix of inter-core communication
The matrix elements are the amount of data transferred across the cores.

### *Algo 4 Steps:*

1. Finds the frequency of communication pattern across
different cores.
2. Analyses the output data set size of divergent functional units of all the cores
3. Relates 1 and 2 to arrive at the amount of data transferred across any two cores.

This algorithm generates the adjacency matrix for the formed cores by analyzing the functional units of the core and assigns respective weight.

The set of algorithms described in the deliver the near optimal count of types of core and their respective near optimal count of cores and the associated inter core communication. This weighted connectivity matrix, describing the inter-core communication pattern is analyzed and partitioned into different weighted sub-matrices. This analysis is based on the dense inter core communication pattern evident from the communication matrices.

# VIII. Inter Core Heterogeneous Network Formation: Given Core Types and Respective Count

As discussed in the algorithm 5 of previous section, the individual entries represent the quantum of data in bytes exchanged across the cores. More on this is presented in the case study section IX. A sample inter core communication pattern is shown in Figs.14a and 14b with arbitrary entries representing the quantum of data in bytes exchanged across the cores during simulation over several hundred thousand cycles. The partitions encompassing cores associated with relatively higher amount of data exchanged in bytes is highlighted and illustrated in Figs14.

| 0 | 150 | 185 | 20 | 0 | 30 |
|---|---|---|---|---|---|
| 150 | 0 | 100 | 20 | 45 | 0 |
| 185 | 100 | 0 | 10 | 15 | 10 |
| 20 | 20 | 10 | 0 | 19 | 15 |
| 0 | 45 | 15 | 19 | 0 | 34 |
| 30 | 0 | 10 | 15 | 34 | 0 |

Fig.14b

The simplest way to arrive at the number of switches in a mesh network and hence the mesh size is given below,
Let the number of bytes through a switch = s
Total number of bytes corresponding to a partition =b
The total number of switches making up the mesh = b/s

## Co-Design Automation tool Framework:

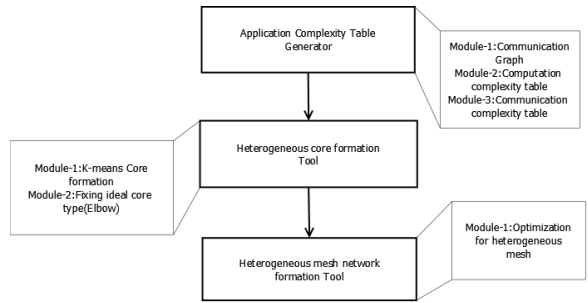

**Fig.15. Co-design automation tool frame work**

The overall flow of the design automation tool is shown in figure 15 based on Fig 12.The simulator and the optimizer tools are presented in [10] with the respective software modules. Based on theapplication complexity modeling presented in [9] a tool has been developed using the set of algorithms whose pseudo-code is construed in the preceding section. Figure 15displays the corresponding software framework.

Description of different modules.

# IX. Co-Design automation: A Case Study

The given multiple applications are shown in the figure for which the heterogeneous many core architecture needs to be generated through co-design automation which involves five phases of development.  The algorithms concerning each of the phases are already presented in section lll

## Step 1-WPPL code for the application

## Step 2: Application Computation Complexity table from the wppl code

| Levels | Computation Complexity | Alfu Units |
|---|---|---|
| 1 | 24,56 | Matmul |
| 2 | 45.56 | Matadd |
| 3 | 64,56 | Mattrans |
| 4 | 56.78 | Matadd2, Matmul2 |
| 5 | 23.67 | Matinv |

Application Communication Complexity Matrix

| Levels | 1 | 2 | 3 | 4 | 5 |
|---|---|---|---|---|---|
| 1 | 0 | 240 | 560 | 456 | 20 |
| 2 | 240 | 0 | 256 | 45 | 0 |
| 3 | 560 | 356 | 0 | 500 | 0 |
| 4 | 456 | 45 | 500 | 0 | 0 |
| 5 | 20 | 0 | 0 | 0 | 0 |

A communication matrix based on model shown in section 3

**Step 3: Types of Core Formation using inter ALFU Communication pattern**

The communication complexity obtained in the previous step level wise can be broken down to get the individual ALFU communication residing in the respective levels. These inter ALFU communication serves as a measure for the core formation, by grouping ALFUs which have more affinity towards each other. ALFU/Scalar inter communication pattern

K-means with number of clusters =6

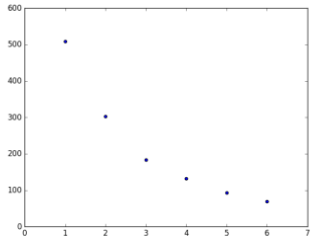

Here we plot the errors on the y –axis and the number of clusters along the x-axis. We can find that the error decreases with the increase in the number of clusters. But choosing a very high number of clusters for the sake of reducing error would not be a good choice. Hence an optimal number of clusters is chosen, where the slope of the curve decreases maximum.

**Step 4: Heterogeneous mesh network for the given the inter core communication pattern across core types**

A mesh network may contain n X m data points connected on a two-dimensional range. Each mesh point will contain buffers and routers to move the data to and fro along the network

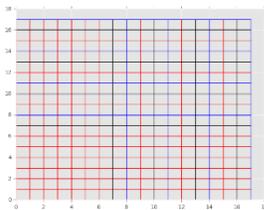

REFERENCES

1. William Rosenbluth,"Design Automation Architecture and Applications",1976, IEEE Computer magazine.
2. Sartajsahni, Atul Bhatt, "THE COMPLEXITY OF DESIGN AUTOMATION PROBLEMS", 1980, ACM 0-89791-020-6/80/0600/0402.
3. Damien Lyonnard, SungjooYoo, AmerBaghdadi,Ahmed A. Jerraya,"Automatic Generation of Application-Specific Architectures for HeterogeneousMultiprocessor System-on-Chip",2001, DAC ACM 1-58113-297-2/01/0006.
4. Rakesh Kumar, Keith I.Farkas, Norman P.Jouppi, ParthasarathyRanganathan, Dean M.Tullsen, "Single ISA Heterogeneous Multi-Core Architectures: The potential for Processor Power Reduction", 2003, Proceedings of the 36th annual IEEE/ACM International Symposium on Micro architecture.
5. Rakesh Kumar, Keith I.Farkas, Norman P.Jouppi, ParthasarathyRanganathan, Dean M.Tullsen, "Single ISA Heterogeneous Multi-Core Architectures for Multithreaded Workload Performance", 2004, Proceedings of the 31st annual internationalsymposium on Computer architecture
6. Rakesh Kumar, Dean M. Tullsen, , Norman P. Jouppi, "Core Architecture Optimization for Heterogeneous ChipMultiprocessors", 2006, PACT
7. Sukhun Kang, Rakesh Kumar, "Magellan: A Search and Machine Learning-based Framework for Fast Multi-core Design Space Exploration and Optimization", 2008, Design Automation and Test.
8. Mark Hammerquist, Roman Lysecky, "DESIGN SPACE EXPLORATION FOR APPLICATION SPECIFIC FPGAS IN SYSTEM-ON-A-CHIP DESIGNS.

WARFT – Waran Research FoundaTion, 45-B, Mahadevan street,West Mambalam, Chennai,India.
Director/ Founder: Prof. N.Venkateswaran

9. "High Performance Computing Co-Design: Application Complexity Design Space, Heterogeneous Many core Architecture Design Space and Parallel Programming Language Design Space",2016,Submitted to IEEE Transactions On Computers for Review.
10. "High Performance Computing Co-Deign: Simulator Design Space and Optimization Design Space", 2016, Submitted to IEEE Transactions On Computers for Review.
11. <<first paper 1>>
12. <<first paper 2>>
13. <<first paper 3>>
14. N.Venkateswaranetal,"On the Concept of Simultaneous Execution of Multiple Applications on Hierarchically Based Cluster and the Silicon Operating System",2008, IEEE IPDPS
15. N.Venkateswaranetal," SCOC IP Cores for Custom Built Supercomputing Nodes", 2012, IEEE Computer Society Annual Symposium on VLSI.

16. **HYPERLINK "http://ieeexplore.ieee.org/search/searchresult.jsp?searchWithin=%22Authors%22:.QT.Benard%20Xypolitidis.QT."**BenardXypolitidis etal, "Towards Architectural Design Space Exploration for Heterogeneous Manycores", 2016, IEEE PDP
17. Yi Zhang etal, "Exploring Many-Core Architecture Design Space for Parallel Discrete Event Simulation", 2014, SIGSIM-PADS'14
18. Amit Kumar Singh etal, "Mapping on Multi/Many-core Systems:Survey of Current and Emerging Trends",2013,DAC
19. climate modeling
21. Aerodynamics ,cfd
22. Brain modeling
23. Design and Analysis of the BlueGene/L Torus Interconnection Network,IBM Report
24. PaulBogdanetal, "Workload Characterization and its impact on Multicore Platform Design", 2010, ACM
25. LINPACK
26. Application cloning 1
27. Application cloning 2
28. Application cloning 3
29. IBS Benchmark, http://web2.clarkson.edu/class/cs644/isolation/index.html
30. Bzip, **www.bzip.org**
31. Jpef, www.jpeg.org
32. Astar, http://web.mit.edu/eranki/www/tutorials/search/
33. LINPACK,"top500.org"
34. Isolation Benchmark Suite, http://web2.clarkson.edu/class/cs644/isolation/index.html
35. JariKreku," Automatic workload generation for system-level exploration based on modified GCC compiler", EDAA ,2010
36. JukkaSaastamoinen,"Application workload model generation methodologies for system-level design exploration",Design and Architectures for Signal and Image Processing (DASIP), 2011 Conference
37. ShekharBorkaretal"On the Role of Co-design in High Performance Computing
38

39 Clay Hughes and Tao LiAccelerating Multi-core Processor Design Space EvaluationUsing Automatic Multi-threaded Workload Synthesis, IEEE, 2008

40. Akash Kumar, "Multiprocessor Systems Synthesis for Multiple Use-Cases of Multiple Applications on FPGA", ACM, 2008

41. ShekarBorkar,"Thousand Cores chips-A technology perspective",DAC'07

42. individual

43. i2

44. i3

45. ShuaiCheetal, "Rodinia: A Benchmark Suite for Heterogeneous Computing",IISWC, 2009

46. J.Cengetal, "MAPS:An integrated Framework for MPSoC Application Parallelization" , DAC 2008.

47. Pluto, "http://www.ece.lsu.edu/jxr/pluto/"

48. CFD,brain modeling
49. " Performance Workload Design", https://www.ibm.com/developerworks/community/.../WorkloadDesign.pdf